\documentstyle[multicol,aps,prl,psfig,amssymb]{revtex}
\begin{document}
\title{Onset of fluidization in vertically shaken granular material}
\author{Thorsten P\"oschel$^{1,2}$, Thomas Schwager$^2$, and Clara 
Salue\~na$^2$}
\address{$^1$ICA 1, Universit\"at Stuttgart, Pfaffenwaldring 27,
D-70569 Stuttgart, Germany\\ 
$^2$Humboldt-Universit\"at zu Berlin,Institut f\"ur Physik,
Invalidenstra\ss e 110, D-10115 Berlin, Germany}
\draft
\date{\today}
\maketitle

\begin{abstract}
  When granular material is shaken vertically one observes convection,
  surface fluidization, spontaneous heap formation and other effects.
  There is a controversial discussion in literature whether there
  exists a threshold for the Froude number $\Gamma=A_0\omega_0^2/g$
  below which these effects cannot be observed anymore. By means of
  theoretical analysis and computer simulation we find that there is
  no such single threshold. Instead we propose a modified criterion
  which coincides with critical Froude number $\Gamma_c=1$ for small
  driving frequency $\omega_0$.
\end{abstract}


\begin{multicols}{2}
\section{Introduction}

When granular material in a rectangular container is imposed to
vertical oscillation under certain conditions one observes a variety
of effects, such as
convection~\cite{EhrichsJaegerKarczmarKnightKupermanNagel:1994,KonvMD,TaguchiPRL},
surface
fluidization~\cite{WarrHuntleyJacques:1995,GallasHerrmannSokolowski:1992FLUI,Taguchi:1992JDP,ClementLudingBlumenRajchenbachDuran:1993,LudingHerrmannBlumen:1994},
spontaneous heap
formation~\cite{Faraday:1831,DouadyFauveLaroche:1989}, surface
patterns~\cite{MeloUmbanhowarSwinney:1994,MetcalfKnightJaeger:1997},
oscillons~\cite{MeloUmbanhowarSwinney:1996} and others. The common
feature of all these effects is that particles change their position
with respect to each other. Provided the particles 
do not change their mechanical properties during observation time
(by polishing, comminution etc.) the condition for this motion is that
neighboring particles separate from each other at least for a small
part of the oscillation cycle $T=2\pi/\omega_0$.

There is a controversial discussion in the literature whether there is
a critical value of Froude number
\begin{equation}
\Gamma_c=A_0\omega_0^2/g\,,
\end{equation}
below which the above mentioned effects vanish, with $A_0$ and
$\omega_0$ being the parameters of the sinusoidal motion of the
container. In {\em many} experimental
observations~(e.g.~\cite{EhrichsJaegerKarczmarKnightKupermanNagel:1994,Taguchi:1992JDP,ClementLudingBlumenRajchenbachDuran:1993,DouadyFauveLaroche:1989,MeloUmbanhowarSwinney:1994,DuranRajchenbachClement:1993,Moreau:1993})
and computer simulations
(e.g.~\cite{Moreau:1993,LudingClementBlumenRajchenbachDuran:1994CONV})
such a critical number $\Gamma_c$ was found. Several authors believe
that the value is $\Gamma_c=1$. In numerical simulations, however,
surface fluidization and convection has been found for $\Gamma
\lesssim
1$~\cite{TaguchiPRL,LudingHerrmannBlumen:1994,BarkerMehta:1992}.
Therefore, some authors believe that $\Gamma$ is not the proper
criterion to determine the degree of fluidization of a granular
system~\cite{GallasHerrmannSokolowski:1992FLUI,BarkerMehta:1993a}.

In this article we discuss the response of granular material with
respect to vertical oscillation in the limit of an one dimensional
approach: the lowest bead of a vertical column of $N$ identical
spherical beads is shaken with $z_0=A_0\cos\omega_0 t$ and the other
beads move due to their interaction force and gravity $g$. We study
the motion of the entire column and can show that particles can lose
contact to their neighbors even for the case that
$\Gamma=A_0\omega_0^2/g$ is significantly less than 1.

Adjacent spheres $k$ and $k+1$ of radius $r$ and mass $m$ at vertical
positions $z_k$ and $z_{k+1}$ interact with their next neighbors by
\begin{equation}
  \label{forcelaw}
  F_{k,k+1}=-\sqrt{r}\left(\mu\xi^{3/2}_{k,k+1}+
\alpha\dot{\xi}_{k,k+1}\sqrt{\xi_{k,k+1}}\right)~~,
\end{equation}
with $\mu$ and $\alpha$ being elastic and dissipative material
constants, i.e. functions of Young modulus, Poisson ratio and
dissipation rate (for details of the derivation of Eq. (\ref{forcelaw})
see~\cite{BSHP}). $\xi$ is the compression $2r-\left|z_k-z_{k+1}\right|$ of
the spheres. The height of the column is $L=2Nr$. Expression
(\ref{forcelaw}) is valid if the typical relative velocities of adjacent
spheres are far below the speed of sound in the material of the
spheres. Certainly this condition holds for typical vibration
experiments.

Introducing new coordinates $u_k=z_k - 2rk$ ($k=0\dots N$) the
compression of two adjacent spheres is
\begin{equation}
  \label{overlap}
  \xi_{k,k+1}=u_k-u_{k+1}\,.
\end{equation}
Applying these definitions in Eq.~(\ref{forcelaw}) and adding gravity
$g$ we get
\begin{eqnarray}
  \label{eqnofmotion}
  \ddot{z}_k&=&\frac{1}{m}\left(F_{k,k+1}-F_{k-1,k}\right)-g
\label{diskret}\\
    F_{k,k+1}&=&-\mu\sqrt{r}\left(u_k-u_{k+1}\right)^{3/2}-  
               \label{eqnofmotion1}
\nonumber\\             &&~~~-
\alpha\sqrt{r}\left(\dot{u}_k-\dot{u}_{k+1}\right)
                 \sqrt{u_k-u_{k+1}}\,.\nonumber
\end{eqnarray}
The $0$-th sphere is fixed at the oscillating table, hence its position is
\begin{displaymath}
  z_0(t)=u_0(t)=A_0\cos\omega_0t.
\end{displaymath}

We are interested in the critical parameters of driving ($A_0$,
 $\omega_0$) when the $N$-th particle loses contact, i.e. when
$u_N>u_{N-1}$. 
We define the ``response''
$R(\omega_0)$ as the ratio $A_N/A_0$ where $A_N$ is the amplitude of
the $N$-th particle at frequency $\omega_0$ and $A_0$ is the amplitude
of the driving vibration. $R(\omega_0)$ can be calculated by
convoluting the motion $z_N(t)$ with $\exp(i\omega_0t)$. Suppose
$A_N\omega_0^2/g \ge 1$ the $N$-th particle separates from the
$N-1$-st. If we would find $A_0<A_N$ the critical Froude number
$\Gamma_c=A_0\omega_0^2/g$ would be less than 1. We will show that
there is a range for $\omega_0$ where this is the case.

In the next section we will formulate the problem in a continuum
approach and derive a nonlinear partial differential equation for the
motion of the column of particles. This equation is solved in section
\ref{sec:elastic} in the limit of elastic material properties, i.e. by
dropping the dissipative terms. Once the solution for the elastic case
has been discussed in detail, it is easier to study the influence of
the dissipative term and to derive the solution of the full equation
of motion which is done in section \ref{sec:full}. Section
\ref{sec:numeric} compares the analytic results with a molecular
dynamics simulation of the original (discrete) problem stated in
Eq. (\ref{diskret}). Finally, we discuss the results.

\section{Continuum approach}

To study the system analytically we use an one dimensional continuum
approach. To this end we perform a Taylor expansion of the force with
respect to the radius $r$ and consequently consider the limit $r\to
0$, $N\to\infty$ with $2rN=L=\mbox{const}$.  First we have to replace
the displacements $u_k$ by $u(2kr)$ introducing the displacement field
$u(z)$ which is a continuous function of $z$. With Eq. (\ref{overlap})
we find from Taylor expansion
\begin{displaymath}
  \xi_{k,k+1}=u(2kr)-u(2kr+2r) 
   = -\left.2r\left(\partial u/\partial z\right)\right|_{z=2kr}\,.
 \end{displaymath}
The net force experienced by the $k$-th particle is
\begin{eqnarray*}
  F_k&=&F_{k,k+1}-F_{k-1,k}\nonumber\\
  &=&-\mu\sqrt{r}\left(\xi^{3/2}_{k,k+1}
    -\xi^{3/2}_{k-1,k}\right)
\nonumber\\  &&
-\alpha\sqrt{r}\left(\dot{\xi}_{k,k+1}\sqrt{\xi_{k,k+1}}
-\dot{\xi}_{k-1,k}\sqrt{\xi_{k-1,k}} \right)\\
  &=&-2\sqrt{2}r^2\mu\left[\left(-\frac{\partial u_k}{\partial z}
\right)^{3/2}-\left(-\frac{\partial u_{k-1}}{\partial z}
\right)^{3/2}\right]\nonumber\\
  &&+2\sqrt{2}r^2\alpha\left[\frac{\partial^2u_k}
{\partial t\partial z}\sqrt{-\frac{\partial u_k}{\partial z}}
-\frac{\partial^2u_{k-1}}{\partial t\partial z}
\sqrt{-\frac{\partial u_{k-1}}{\partial z}}\right]\,,
\end{eqnarray*}
with the abbreviations 
\begin{eqnarray}
u_k &=& u(2kr)\\
\frac{\partial u_k}{\partial z}&=&\left.\frac{\partial u}
{\partial z}\right|_{z=2kr}\,.
\end{eqnarray}
Both expressions in
square brackets are expanded again and Eq.~(\ref{eqnofmotion1}) 
turns into
\begin{displaymath}
  \frac{F_k}{m}=\frac{3\sqrt2}{\pi\rho}\left[-\mu\left(
-\frac{\partial u}{\partial z}\right)^{3/2}+
  \alpha\frac{\partial^2u}{\partial t \partial z}
      \sqrt{-\frac{\partial u}{\partial z}}\right]\,.
\end{displaymath}
With 
\begin{displaymath}
\kappa=\frac{3\sqrt2\mu}{\pi} \rho\,,~~ \beta=\frac{3\sqrt2\alpha}{\pi\rho}
\end{displaymath}
the continuum formulation of (\ref{diskret}) is
\begin{eqnarray}
  \label{elastic}
&& \frac{\partial^2u}{\partial t^2}=-g-\frac{\partial}{\partial z}
  \left[\kappa\left(-\frac{\partial u}{\partial z}
\right)^{3/2}-\beta\frac{\partial^2u}
{\partial t\partial z}\sqrt{-\frac{\partial u}{\partial z}}\right]\,,\\
&& \left.\frac{\partial u}{\partial z}\right|_{z=L} = 0\nonumber
\end{eqnarray}
where $g$ accounts for the gravitational force. 

\section{Limit of elastic particles}
\label{sec:elastic}

In the following we consider
Eq. (\ref{elastic}) the limit of no damping ($\beta=0$). Using new
variables
\begin{eqnarray*}
x=1-\frac{z}{L}\,&,&~~~\tau=\left(\frac{g\kappa^2}{L^5}\right)^{1/6}t\,\\
\Omega=\left(\frac{L^5}{g\kappa^2}\right)^{1/6}\omega\,&,&
~~~\gamma=\left(\frac{g^2L^5}{\kappa^2}\right)^{1/6}  
\end{eqnarray*}
Eq.~(\ref{elastic}) turns into 
\begin{eqnarray}
&& \frac{\partial^2 u}{\partial \tau^2} = -\gamma^2+
 \frac{1}{\gamma}\frac{\partial}{\partial x}
 \left[\left(\frac{\partial u}{\partial x}\right)^{3/2}\right]
\label{undamped}\\
&& \left.\frac{\partial u}{\partial x}\right|_{x=0} = 0 \,.
\label{boundary}
\end{eqnarray}
Equation~(\ref{undamped}) is defined in the range $x\in[0,1]$. The time
independent solution $U(x)$ of (\ref{undamped}) is
\begin{equation}
  \label{static}
  U(x)=\frac{3}{5}\gamma^2\left(x^{5/3}-1\right)\,.
\end{equation}
The solution of (\ref{undamped}) can be considered as a superposition
of the static solution (\ref{static}) and a perturbation $w(x,\tau)$.
Inserting $u=U+w$ in (\ref{undamped}) we find:
\begin{eqnarray}
  \frac{\partial^2 w}{\partial \tau^2} 
   &=&-\gamma^2+\frac{1}{\gamma}\frac{\partial}{\partial x}
     \left[\frac{\partial U}{\partial x} + 
      \frac{\partial w}{\partial x}\right]^{3/2}\nonumber \\
   &\approx&-\gamma^2+\frac{1}{\gamma}\frac{\partial}{\partial x}
      \left[\left(\frac{\partial U}{\partial x}\right)^{3/2}+
      \frac{3}{2}\sqrt{\frac{\partial U}{\partial x}} 
      \frac{\partial w}{\partial x}\right]\nonumber\\
   &=&\frac{3}{2}\frac{\partial}{\partial x}
      \left[x^\frac{1}{3}\frac{\partial w}{\partial x}\right]
  \label{wave}
\end{eqnarray}
By separation of variables $w =T(\tau,\Omega) X(x,\Omega)$, i.e. a
standing wave Ansatz we obtain two ordinary differential equations for
$T$ and $x$:
\begin{eqnarray}
  \ddot{T}&=&-\Omega^2T\\ 
  \frac{3}{2}\frac{d}{dx}
  \left(x^{1/3}\frac{dX}{dx}\right)&=&-\Omega^2X\,\label{eq:xxx},
\end{eqnarray}
with $\Omega$ being a real number. For $T(\tau,\Omega)$ one
gets
\begin{displaymath}
  T\sim\exp(i\Omega\tau)\,.
\end{displaymath}
The solution of the spatial equation (\ref{eq:xxx})
can be found using the Ansatz
\begin{displaymath}
  X(x,\Omega)=x^{1/3}f(y),~~~~~
  y=\frac{2}{5}\sqrt{6}\Omega x^{5/6}\,,
\end{displaymath}
which yields
\begin{equation}
  y^2\frac{d^2f}{dy^2}+y\frac{df}{dy}+\left(y^2-\frac{4}{25}\right)f=0\,.
\label{Bessel}
\end{equation}
Eq.~(\ref{Bessel}) is the Bessel equation of order $2/5$. Hence the
solution of (\ref{eq:xxx}) is
\begin{equation}
  X(x,\Omega)=\left(\frac{6}{25}\right)^{1/5}
  \Gamma\left(\frac{3}{5}\right)\Omega^{2/5} x^{1/3}
  J_{-\frac{2}{5}}\left(\frac{2}{5}\sqrt{6}\Omega x^{5/6}\right)\,.
\label{Bessel.ort}
\end{equation}
An expression containing $J_{2/5}$ would be a solution too, however,
it does not satisfy the condition (\ref{boundary}). The prefactor in
(\ref{Bessel.ort}) has been chosen to assure $X(0,\Omega)=1$.

Hence the solution for a single vibrational mode $u_\Omega$ is
\begin{equation}
  \label{eq:singlemode}
  u_\Omega=\exp\left(i\Omega\tau\right)X(x,\Omega)\,.
\end{equation}

Without prior knowledge the full solution of Eq. (\ref{wave}) has to 
be assumed to be a superposition of vibrational modes for all real 
(rescaled) frequencies $\Omega$:
\begin{equation}
  u=\int_{-\infty}^{\infty}d\Omega\,A(\Omega)
  \exp\left(i\Omega\tau\right)X(x,\Omega)\,.
\label{eq:13}
\end{equation}
In the steady state of pure sinusoidal excitation of the base, i.e.,
when all non-oscillatory perturbations which originate from the
initialization have been damped out, Eq. (\ref{eq:13}) is the full
(steady state-) solution of Eq. (\ref{wave}).

The function $A(\Omega)$ represents the excitation of the mode at
frequency $\Omega$. The boundary condition at the top of the chain is
automatically satisfied, whereas the boundary condition at the bottom
reads
\begin{eqnarray}
  u(1,\tau)&=&\int_{-\infty}^{\infty}d\Omega\,A(\Omega)
  \exp\left(i\Omega\tau\right)X(1,\Omega)\label{eq:enforce_bc1}\\
  &=&A_0\cos\Omega_0\tau\,.
\end{eqnarray}
One can see that the integrand of Eq.~(\ref{eq:enforce_bc1}) can be
nonzero only for $\Omega\ne\Omega_0$. This means that for
$\Omega\ne\Omega_0$ either $A(\Omega)$ or $X(1,\Omega)$ have to be
zero, i.e. for all frequencies for which $X(1,\Omega)$ is nonzero the
respective amplitude must be zero, whereas for all frequencies which
are a root of $X(1,\Omega)=0$ the amplitude can be nonzero. Therefore,
we find that the full solution of Eq.~(\ref{wave}) is a superposition
of the vibrational mode of the frequency of shaking $\Omega_0$ and of
a discrete set of frequencies $\Omega_k$ ($k=1\dots\infty$).

Note that $\Omega_k$ are no rational multiples of each others since
the roots of Bessel-functions are incommensurable (see
Eq.~(\ref{Bessel.ort})).  Therefore, to determine the maximum
acceleration of the topmost particle it is sufficient to consider only
the mode of the external excitation. All other vibrational modes can
only further increase the maximal acceleration.

\medskip
The above defined response $R$ is the ratio $A_N/A_0$. Since the
zeroth particle corresponds to $x=1$ and the N-th to $x=0$ we can
write
\begin{eqnarray}
  R^{-1}(\Omega_0)&=&\left|\frac{X(1,\Omega_0)}{X(0,\Omega_0)}\right| = 
  \left|X(1,\Omega_0)\right|\nonumber \\
  &=&\left(\frac{6}{25}\right)^{1/5}
      \Gamma\left(\frac{3}{5}\right)\Omega_0^{2/5}\left|J_{-\frac{2}{5}}
      \left(\frac{2}{5}\sqrt{6}\Omega_0 \right)\right|\,.
  \label{response}
\end{eqnarray}
The response $R$ is an amplification factor, hence, the value
$g/R(\Omega_0)$ is the critical acceleration of the driving vibration 
\cite{NoHigherOrders}. 
$R$ is larger than 1 for all driving frequencies $\omega_0$. This
means that for {\em any} driving frequency $\omega_0$ and driving
amplitude $A_0$ the amplitude of the top particle of the column $A_N$
at frequency $\omega_0$ will be larger than $A_0$. Therefore, for $A_N
\omega_0^2/g=1$, i.e. when the $N$-th particle separates from the
$N-1$-st, we find $A_0\omega_0^2/g = \Gamma_c < 1$.
 
According to the above arguments we have to replace the condition
$\Gamma\ge 1$ which was supposed to be the precondition for surface
fluidization, convection etc., by
\begin{equation}
  \label{result}
  A_0\omega_0^2/g = \Gamma \ge R^{-1}\left(\omega_0\right).
\end{equation}

The function $R^{-1}(\omega_0)$ over $\omega_0$ is drawn in
Fig.~\ref{fig:response} (dash-dotted line, $R^{-1}_{el}$). For the
system parameters we used $A_0=0.01\,{\rm mm}$, elastic constant
$\kappa=2.8\cdot 10^4\,{\rm m}^{2}/{\rm sec}^2$ (rubber with Young
modulus $Y=4\cdot 10^7\,{\rm Pa}$) and $L=0.6\,{\rm m}$. The curve
reveals pronounced resonances at Eigenfrequencies $\omega_k$ where
$R^{-1}$ becomes minimal (only the first resonance is shown in
Fig.~\ref{fig:response}).

All experiments on surface fluidization and convection which can be
found in literature have been performed far below the first
resonance. Therefore, of particular interest to practical purposes is
the limit of small frequency $\omega_0$, i.e. below the first
Eigenfrequency. The Taylor expansion of Eq.~(\ref{response}) yields
$R^{-1}(\Omega_0)$ for small $\Omega_0$
\begin{eqnarray}
  R^{-1}&=&1-\frac{2}{5}\Omega_0^2+{\cal O}(\Omega_0^4)\nonumber \\
  &=&1-\frac{2}{5}\left(\frac{L^5}{g\kappa^2}\right)^{1/3}\omega_0^2+
       {\cal O}(\omega_0^4)\,.
\label{Taylor.elas}
\end{eqnarray}

Given the container vibrates with frequency $\omega_0$. Then for the
critical amplitude $A_0$ of the vibration when the top particle
separates, i.e. when the material starts to fluidize, one finds
\begin{equation}
  A_0=\frac{g}{\omega_0^2}-\frac{2}{5}
       \left(\frac{L^5}{g\kappa^2}\right)^{1/3}\,.
\end{equation}
\begin{minipage}{8.5cm}
\begin{figure}[htb]
  \centerline{\psfig{figure=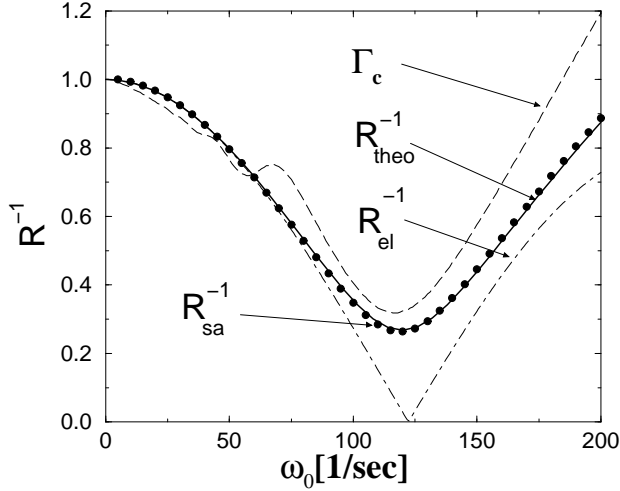,width=8cm}}
  \caption{The response function $R_{\rm el}^{-1}$ due to
  Eq.~(\ref{response}) for a column of elastic particles
  (dash-dotted). For dissipative particles ($\beta=$127m$^2$/sec):
  circles, $R_{\rm sa}^{-1}$: numerical integration of
  Eq.~(\ref{eqnofmotion}) at small amplitude; full line,
  $R_{\rm theo}^{-1}$: analytical solution (\ref{RofBeta}) of the full
  Eq.~(\ref{elastic}) including dissipation; dashed line,
  $\Gamma_c(\omega_0)$: result of a direct simulation of
  Eq. (\ref{eqnofmotion}) (explanation see text)}
\vspace{0.5cm}
  \label{fig:response}
\end{figure}
\end{minipage}
Surprisingly even for very small frequencies where $R^{-1}\to 1$ one
finds that the critical amplitude is reduced by a constant as compared
with $g/\omega_0^2$. So although the value of the response function 
comes arbitrarily close to one, the critical {\em amplitude} differs 
from the expected one by a constant. However, this does not mean that 
the critical Froude number becomes a constant. 

Eq.~(\ref{Bessel.ort}) describes the behavior of the column of grains
for the case of purely elastic contact ($\alpha=0$). If the
dissipative material properties are taken into consideration the full
equation (\ref{elastic}) has to be solved which will be discussed in
the following section.

\section{Dissipative particle interaction}
\label{sec:full}

We will consider, as before, small perturbations $w$ about the static
deformation of the chain under gravity which propagate from the
bottom. The dissipative term is characterized by the parameter $\beta$
in Eq.~(\ref{elastic}). From this equation, again introducing the
static solution given by Eq.~(\ref{static}) and by using the same
transformation for the spatial coordinate $x \equiv 1-z/L$, one
obtains the corresponding linearized wave equation for dissipative
materials,
\begin{equation}
\frac{\partial^{2} w}{\partial t^{2}} = \left(\frac{g}{\kappa
L^5}\right)^{\!\frac{1}{3}} \left[\frac{3}{2}\, \kappa \,
\frac{\partial}{\partial x} \left(\!x^{\frac{1}{3}} \frac{\partial
w}{\partial x} \right) + \beta \, \frac{\partial}{\partial
x}\left(\!x^{\frac{1}{3}} \frac {\partial^2 w}{\partial t \partial x}
\right)\right].
\label{beforeF}
\end{equation}
In this case it is less useful to introduce the rescaled time variable
$\tau$, while it proves convenient to define
\begin{equation}
\left(\frac{\kappa L^5}{g}\right)^{1/6} \equiv \ell\,,
\end{equation}
being the natural length scale coming out from the analysis. 
By means of the Fourier transform,
\begin{equation}
{\mathcal W} (x,\omega) = \frac{1}{\sqrt{2\pi}} \int_{-\infty}^\infty
e^{-i\omega t} \, w (x,t)\, dt,
\end{equation}
equation (\ref{beforeF}) becomes
\begin{equation}
-\omega^2 {\mathcal W} = \frac{\frac{3}{2} \kappa -i \omega
\beta}{\ell^2}\, \frac {\partial}{\partial x} \left(x^{1/3}
\frac{\partial {\mathcal W}}{\partial x} \right) \, ,
\label{afterF}
\end{equation}
which has the same structure as Eq.~(\ref{eq:xxx}). Hence the same
transformations apply in this case and the general
solution reads finally
\begin{eqnarray}
{\mathcal W}(x,\omega) &=& x^{1/3} \left[ C_1 J_{\frac{2}{5}} \left(
\frac{\frac{6}{5}\omega \ell}{\sqrt{\frac{3}{2}\kappa -i \omega
\beta}}x^{5/6}\right) +
\right.\nonumber\\ &&~~+\left. 
C_2
J_{-\frac{2}{5}} \left( \frac{\frac{6}{5} \omega
\ell}{\sqrt{\frac{3}{2} \kappa -i \omega \beta}}x^{5/6}\right)
\right].
\end{eqnarray}
The part of the solution depending on $J_{2/5}$ carries a divergence
at $x=0$ ($z=L$), and therefore $C_1=0$ is required for the solution
to be physical. The condition of the free end at $x=0$ ($z=L$) is 
satisfied
automatically, as in the case $\beta = 0$. The solution has exactly
the same structure as the solution of the elastic problem,
Eq. (\ref{Bessel.ort}), and the only change is that the argument of
the Bessel function has an imaginary part. If one considers only the
mode $\omega_0$, which propagates from the bottom ($x=1$) with
amplitude $A_0$, the solution reads
\begin{equation} 
w(x,t) = \mbox{\rm Re} \left\{ A_0 \, e^{i\omega_0 t}
\,x^{\frac{1}{3}} \,\frac{ J_{-\frac{2}{5}}
\left(\displaystyle\frac{\frac{6}{5}\omega_0 \ell}{\sqrt{\frac{3}{2}
\kappa -i \omega_0 \beta}}~x^{5/6}\right) } {J_{-\frac{2}{5}}
\left(\displaystyle\frac{\frac{6}{5} \omega_0
\ell}{\sqrt{\frac{3}{2}\kappa -i \omega_0 \beta}}\right)}\right\}\,.
\end{equation}

The fluidization condition at the top of the chain
$|\partial^2w/\partial t^2(x=0,t)| > g$ can be written in general
\begin{eqnarray} 
&&R^{-1}(\omega_0)\nonumber\\
&&~~\equiv \Gamma \left(\frac{3}{5}\right)
\left|
\left(
\frac{\frac{3}{5}\omega_0 \ell}{ \sqrt{\frac{3}{2}\kappa -i \omega_0 
\beta}}\right)^{2/5}
J_{-\frac{2}{5}} 
\left( 
\frac{\frac{6}{5} \omega_0 \ell}{\sqrt{\frac{3}{2} \kappa -i \omega_0 
\beta}} 
\right)
\right|
\nonumber\\
&&~~< 
\frac{A_0\,\omega_0^2}{g}~.
\label{RofBeta}
\end{eqnarray}
Since the Bessel functions of the first class only have zeros on the
real axis, $R^{-1}(\omega_0)$ can no longer be zero for any frequency
if $\beta \neq 0$. This means that the sharp resonances displayed by
$R(\omega_0)$ when $\beta=0$ disappear and are replaced by more or
less pronounced minima in dependence on the damping constant $\beta$.
This can also be observed in Figure \ref{fig:response} (full
line). Increasing values of $\beta$ make the response smoother,
deviating from that of the elastic case earlier, and the local minima
translate on the frequency axis appreciably, see Figure
\ref{fig:damping}.
\begin{minipage}{8.5cm}
\begin{figure}[htbp]
 \centerline{\psfig{figure=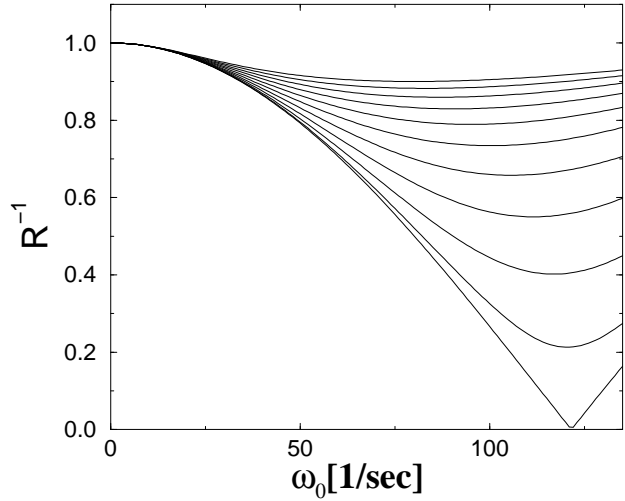,width=8cm}}
 \caption{The response function over the frequency for different
 dissipative constants: $\beta=$(0, 100, 200, \dots , 1000) m$^2$/sec
 (bottom to top). With increasing damping the minimum becomes less
 pronounced and shifts to lower frequencies.}
  \label{fig:damping}
\vspace{0.5cm}
\end{figure}
\end{minipage}

Analogue to the elastic case (Eq.~(\ref{Taylor.elas})), for small
frequencies $R^{-1}(\omega_0)$ can be expanded into a Taylor series:
\begin{eqnarray} 
\label{Taylor.diss}
R^{-1} &\simeq& 1 - \omega_0^2\,\frac{2}{5}\left(
\frac{L^5}{g\kappa^2} \right)^{1/3} 
\\ &&
+ \omega_0^4\left(
\frac{L^5}{g\kappa ^2} \right)^{2/3} \left( \frac{3}{100} +
\frac{8}{45} \beta^2 \left( \frac{g}{\kappa^4L^5} \right)^{1/3}\right)
\, .
\end{eqnarray}
The contribution due to the dissipative parameter enters the Taylor
expansion at the fourth power of the frequency. Therefore, the
analysis of the elastic case given by Eq.~(\ref{Taylor.elas}) remains
valid for small frequencies.

It is interesting to note that due to Eq. (\ref{Taylor.diss}) there
{\em always} exists a global minimum below the value
$R^{-1}(\omega_0)=1$, regardless of the value of the dissipative
parameter. We want to study for which range of frequencies the inverse
response $R^{-1}$ is smaller than one for varying damping constant
$\beta$: The lower boundary of this interval obviously is $\omega=0$.
To determine the upper boundary $\omega_{\rm max}$ we solved
numerically the equation $R^{-1}(\omega_{\rm max})=1$ for different
values of $\beta$. The result of this calculation is shown in Fig.
\ref{fig:omega1}.
\begin{minipage}{8.5cm}
\begin{figure}[htbp]
    \centerline{\psfig{figure=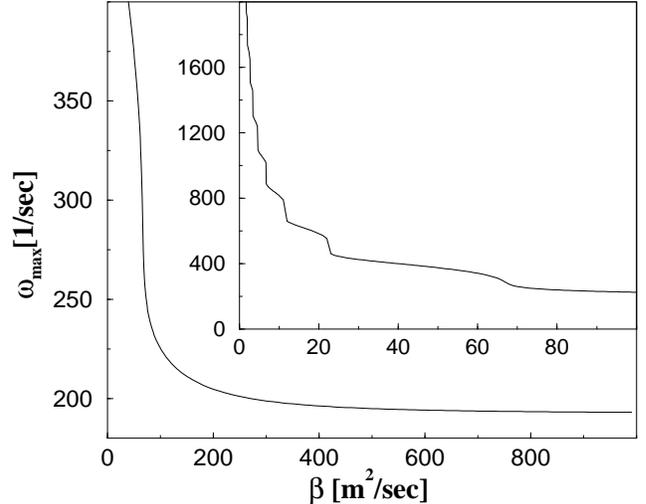,width=7.8cm}}
\vspace{0.4cm}
  \caption{The frequency $\omega_{\rm max}$ at which the inverse
  response function becomes larger than one is almost constant with
  varying dissipative parameter $\beta$ as long as the damping is not
  to small. For very low dissipative parameter $\beta$ one finds a
  non-smooth function (small figure).}
  \label{fig:omega1}
\vspace{0.2cm}
\end{figure}
\end{minipage}

One can see that for high enough damping this frequency $\omega_{\rm
max}$ varies only slowly with $\beta$.  The curve almost saturates at
$\omega_{\rm max}\approx 200\,{\rm sec}^{-1}$ which is close to the
first maximum of the undamped inverse response (see
Fig. \ref{fig:steps}, dash-dotted line). Although the valley of the
inverse response function becomes smaller with increasing damping
(Fig. \ref{fig:omega1}), even for larger damping $R^{-1}$ is smaller
than one in a finite frequency interval, i.e. the effect of amplitude
amplification exists for almost the entire range of frequencies
between zero and the first maximum of the undamped inverse response.
\begin{minipage}{8.5cm}
\begin{figure}[htbp]
    \centerline{\psfig{figure=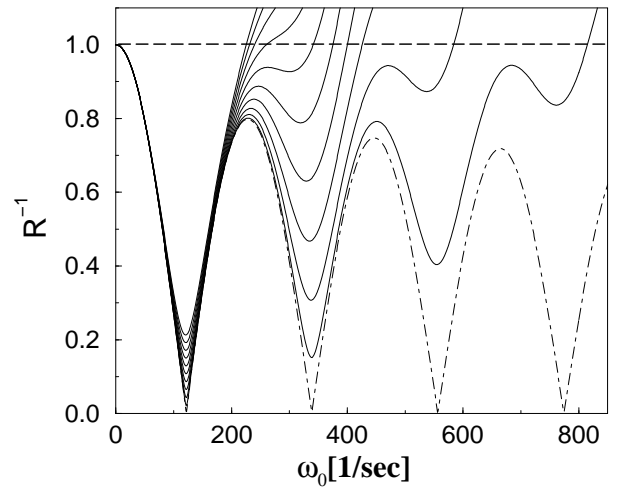,width=7.8cm}}
  \caption{The response function for different values of the
  dissipative constant $\beta=$10, 20, 30,\dots,100\,m$^2$/sec (bottom to top)
  together with the elastic curve $\beta=0$ (dash-dotted). With
  decreasing $\beta$ the curves are influenced by higher order minima
  of the response function. This explains the steps in the curve drawn
  in Fig.~\ref{fig:omega1} for small values of the dissipation
  $\beta$.}
  \label{fig:steps}
\vspace{0.2cm}
\end{figure}
\end{minipage}

For small enough damping the frequency range of amplitude
amplification $R^{-1}\left(\omega_0\right)<1$ extends beyond the
position of the first maximum of the undamped inverse response. Figure
\ref{fig:steps} shows the response function for different dissipative
parameter $\beta$ together with the elastic case ($\beta=0$,
dash-dotted). The frequency $\omega_0$ at which $R^{-1}(\beta)=1$
increases with decreasing damping. 
In this Fig. \ref{fig:steps} one can clearly see the extension of the 
range of amplitude amplification. 
If we decrease the damping parameter, starting at high values where the 
amplifying range is limited to the first ``well'' we cannot expect a
large change since, as long as we remain limited to the first well, there 
is a upper bound (the frequency of the first maximum of $R^{-1}$) to this 
range. Reducing $\beta$ further we will eventually reach values for which 
the amplifying range extends to the second well. Even for values of
$\beta$ which are only slightly below this threshold the range will now 
span again almost the entire range of the second well up to the frequency 
of the second maximum. So there will be rather sharp steps in the 
dependence of the upper range limit of the damping instead of only gradual
changes. This behaviour explains the steps in the closeup of 
Fig. \ref{fig:omega1}.

\section{Numerical results}
\label{sec:numeric}

To check the analytical results and in particular the validity of the
continuum approach we calculated $R^{-1}$ for a finite value of
damping $\beta$ from the numerical simulation of
Eq.~(\ref{eqnofmotion}). The circles in Fig.~\ref{fig:response}
display the reciprocal response $R^{-1}$ over $\omega_0$ with fixed
amplitude $A_0=0.01\,{\rm mm}$, elastic constant $\kappa=2.8\cdot
10^4\,{\rm m}^{2}/{\rm sec}^2$ (rubber with Young modulus $Y=4\cdot
10^7\,{\rm Pa}$) and $L=0.6\,{\rm m}$.  Fig.~\ref{fig:response} shows
that for small frequency $\omega_0$ and small damping $\alpha$ the
undamped theoretical curve (dash-dotted) agrees well with numerical
data. If one compares the numerical result with the damped solution
according to Eq. (\ref{RofBeta}) (full line in
Fig. \ref{fig:response}) the agreement with theory is very well.

To check the validity of our linear theory we also determined directly
by integrating Eq.~(\ref{eqnofmotion}) at which Froude number
$\Gamma_c$ the particles start to jump. The results of this
calculation are shown in the dashed curve in Fig.~\ref{fig:response}
and agree well with the linear theory. To obtain the value of
$\Gamma_c$ for a given frequency $\omega_0$ we determined an upper
bound $A_0^{+}$ for the critical amplitude where one observes jumping
and a lower bound $A_0^{-}$ where no jumping occurs. Then we narrowed
the interval $A_0^{+}-A_0^{-}$ by testing an amplitude between
$A_0^{+}$ and $A_0^{-}$ until $(A_0^{+}-A_0^{-})/A_0^{+} < 10^{-3}$.

\section{Discussion}

For the case of a vertical column of viscoelastic spheres we derived a
linear wave equation in one dimensional approximation. We have shown
that the sphere on top of the column $N$ can separate from the
$N-1$-st even if the container is oscillated with $A_0\omega^2_0/g <
1$.  As main result we derived a modified condition for the topmost
particle to separate from its neighbor. We could show that instead of
the widely accepted condition $\Gamma_c\equiv A_0\omega_0^2/g>1$ one
has to satisfy $A_0\omega_0^2/g>R^{-1}$ where $R^{-1}$ is a function
of $\omega_0$. We have shown that independent on the material
properties there exists {\em always} a range
$\omega_0\in[0..\omega_{\rm max}]$ for which the amplitude of
vibration $A_0$ is amplified, i.e. for which the top particle can
separate (the material fluidizes) even if $A_0\omega_0^2/g <1$.
Numerical calculations agree well with the analytic results.

Whereas the critical Froude number $\Gamma_c\ge 1$ is certainly the
proper criterion to predict whether a single rigid particle will jump
on a vibrating table, we suspect that this number is not suited to
describe the behavior of a vibrated column of spheres, and the more it
is not a criterion for surface fluidization of a three dimensional
granular material.

The authors wish to thank E.~Cl\'ement, N.~Gray, H.~J.~Herrmann,
H.~M.~Jaeger, S.~Luding, S.~Roux and L.~Schimansky-Geier for helpful
discussion.

\end{multicols}

\begin{references}
\bibitem{NoHigherOrders} Here we consider only the mode of external
excitation $\omega_0$. Since the higher frequencies $\omega_k$ are
noncommensurable with $\omega_0$ and with each others the maximum
acceleration of the superposition must be higher than
$R(\omega_0)A_0\omega_0^2$. Hence consideration of higher modes will
further increase the effect discussed here.
\end{references}
\end{document}